\documentclass[ reprint, amsmath,amssymb, aps,prl,superscriptaddress]{revtex4-2}

\usepackage{graphicx}
\usepackage{dcolumn}
\usepackage{bm}

\makeatletter
\newcommand*{\balancecolsandclearpage}{%
  \close@column@grid
  \clearpage
  \twocolumngrid
}
\makeatother

\begin{document}

\title{Observing the Glass and Jamming Transitions of Dense Granular Material in Microgravity}

\author{Christopher Mayo}\affiliation{Institute for Materials Physics in Space, German Aerospace Center (DLR), 51170 Cologne, Germany}

\author{Marlo Kunzner}\affiliation{Institute for Materials Physics in Space, German Aerospace Center (DLR), 51170 Cologne, Germany}

\author{Matthias Sperl}\email{Matthias.Sperl@dlr.de}\affiliation{Institute for Materials Physics in Space, German Aerospace Center (DLR), 51170 Cologne, Germany}\affiliation{Department for Theoretical Physics, University of Cologne, 50937 Cologne, Germany}

\author{Jan Philipp Gabriel}\email{Jan.Gabriel@dlr.de}\affiliation{Institute for Materials Physics in Space, German Aerospace Center (DLR), 51170 Cologne, Germany}
   

\begin{abstract}  
The present study investigates a weakly pulsed granular system of polystyrene spheres under long-time microgravity conditions on the International Space Station (ISS). The spheres are measured using Diffusing Wave Spectroscopy (DWS) and are described by mean square displacements (MSDs). Our aim is to use this technique to show the first experimental evidence of glassy dynamics in dense granular media in microgravity and subsequently compare these results with ground-based measurements to see how the nature of these dynamics change without the influence of gravity. Our results show that as we densify the sample in microgravity, glassy dynamics appear at a volume fraction 1.6\% lower than on ground. We also show how the influence of gravity can affect how dense a granular system one can prepare by comparing the final jamming point of our sample on the ISS compared to our ground setup. We show that jamming occurs at a  volume fraction 0.5\% lower in space compared to on ground. Showing that we can create denser states when a granular system is in the presence of a stronger gravitational field. 
\end{abstract}

\maketitle

Granular material is one of the most abundant and widely utilized types of materials in our society \cite{andreotti2013granular,duran2012sands}. Grains feature in every aspect of life, from the coffee one drinks to the paths one walks on \cite{andreotti2013granular,duran1997sables}. They also feature heavily in industries such as pharmaceuticals, agriculture, and energy production \cite{andreotti2013granular}. Understanding the way grains move and interact then becomes a field of key interest, especially when produced, stored and used in a dense state \cite{janssen1895versuche,windows2019janssen}. One of the key characteristics of granular material is gravitational settling \cite{andreotti2013granular,wentworth1922scale}. Gravity is an attractive force which influences the dynamics of the system \cite{rosato2020segregation}. It is then a natural question to ask, what if the force of gravity was removed from such a system? How would the dynamics change? Such a system has been simulated \cite{Opsommer2011} and granular material has been investigated in microgravity experiments prior to this experiment but to this point not in dense states. The previous works have tended to revolve around less dense systems such as granular gases \cite{Evesque,Wang,Piti2022} due to the natural difficulty of capturing the complete dynamics of a dense, opaque system combined with the challenge of obtaining enough experiment time in microgravity to resolve slow relaxation dynamics. This provides the motivation for our investigation; where we used an established light scattering technique such as diffusive wave spectroscopy (DWS) to measure dynamical responses of dense granular systems on the international space station and compare our findings to those seen on ground \cite{born2021soft,weitz1993diffusing,Brown1993Dynamic,Kunzner2025VC}.

Dynamics is a broadly used term to describe how the system moves. The goal of our experiment campaign was to measure the slowing down of dynamics as we densified our granular system to investigate the glass and jamming transition. When considering the state diagram of granular media \cite{trappe2001jamming} one can see that the dynamics are controlled by the density and the driving force. It has long been hypothesized that changing the density of a dynamical granular system, while keeping the driving force constant, should yield similar characteristics of that of a glass\cite{liu1998jamming,berthier2009glasses}. We then looked to measure and quantify the slowing down of the dynamics of our system as we increase our density, to identify the point at which a similar dynamical responses to that of a glass\cite{liu1998jamming} is seen. Measurements were conducted in both microgravity and on ground to enable a direct comparison of the effect gravity has on these transitions. We also provide a basis for the general behavior of dense granular material in microgravity environments which provides insights regarding building material when considering construction on say a lunar-based habitat, where the only source of natural building material is in fact grains \cite{happel1993indigenous}.  

\begin{figure}[t!]
    \centering
    \includegraphics[width=\linewidth]{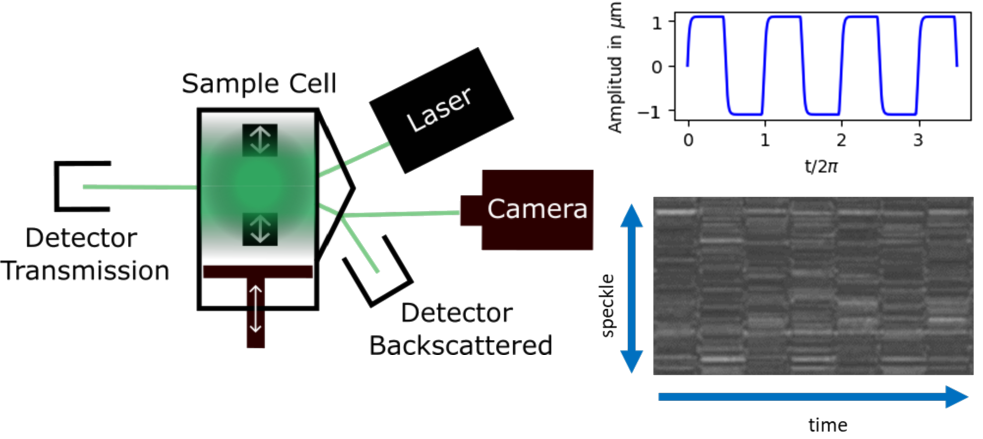}
    \caption{Schematic of the set-up with piezo agitation source in the sample volume with agitation profile, piston for volume control, laser source, detectors for collecting intensity fluctuations in transmission and backscattering geometry, and, camera for time-resolved speckle detection.
    }
    \label{fig1a}
\end{figure}

For our experiment, as shown in Fig.\ref{fig1a}, samples are illuminated with a 532nm laser, and then intensity traces are collected in both the transmission and backscattering geometries. Which we will refer to as 'bulk' and 'wall' dynamics respectfully. The intensity signal is then split and fed into two avalanche photodiodes. These are then cross-correlated using an ALV USB correlator series-7004. Also in the backscattering geometry, is a line camera which records a time-lapse of 500 speckles merged to form a 2D image.
The 140 micron polystyrene spheres are contained in a 11x10x5mm sample cell. The height of the sample cell is adjustable, from 11mm to 3.6mm, which enables our change in volume fraction (VF) \cite{cumberland1987packing}. Where the VF=$V_{Sample}/V_{Sample Cell}$ expressed as a percentage.

The volume of the sample is calculated using the stated density of 1057 $kg/m^3$ of the PS grains and the mass \cite{PhysRevFluids.6.L062301}. The height of the sample cell is given by an encoder position provided by a magnetic encoder scale. Inside the sample cell are 4 piezo cubes, of size 2x2x2mm, which provide an agitation amplitude of 2.2$\mu$m and an acceleration of approximately 1g at 60Hz. The volume of the piezo elements is then deducted from the volume of the container when calculating the VF\cite{cumberland1987packing}. The agitation frequency can be tuned to values between 5Hz and 2kHz. An agitation frequency of 60Hz was chosen after preliminary tests. Two sets of experimental runs can be described, the first is where the VF is increased, this will be referred to as densification experimental runs. Also conducted were expansion runs where the packing fraction was decreased. Measurements were taken at fixed VFs after sufficient equilibrium of 1 hour after each measurement. Each measurement produces an intensity correlation function $g_2$, from which the field correlation, $g_1$, can be calculated via the Siegert-Relation \cite{berne2000dynamic}.
\begin{equation}
    \label{eq:Siegert}
    g_2(t) = 1 + \Lambda\, |g_1(t)|^2
\end{equation}
Where $\Lambda$, refereed to as the coherence area factor which is assumed to be 1. The later analysis assumes that we can connect $g_1(t)$ with the mean square displacement (MSD) $\langle \Delta r^2 \rangle$ by using the DWS approximations \cite{Brown1993Dynamic,weitz1993diffusing}.
\begin{equation}
    \label{eq:FieldAutoCorr}
    g_1(t) = \exp(-\frac{1}{3}\left(\frac{kL}{l^*}\right)^{2} \langle r^2 \rangle)
\end{equation}
where $k=2\pi/\lambda$ is the wave vector and L is the length of the sample cell. The effective mean free path length $l^* = l/(1-<cos(\theta)>)$ where $l$ is the mean free path length determined by the scattering cross-section $\sigma(\theta)$ for more details see \cite{Kunzner2025VC}. A full description of how the oscillation from the piezos influences the correlation function $g_1$ can be seen in the appendix.


\begin{figure*}[th!]
    \centering
    \includegraphics[width=0.95\linewidth]{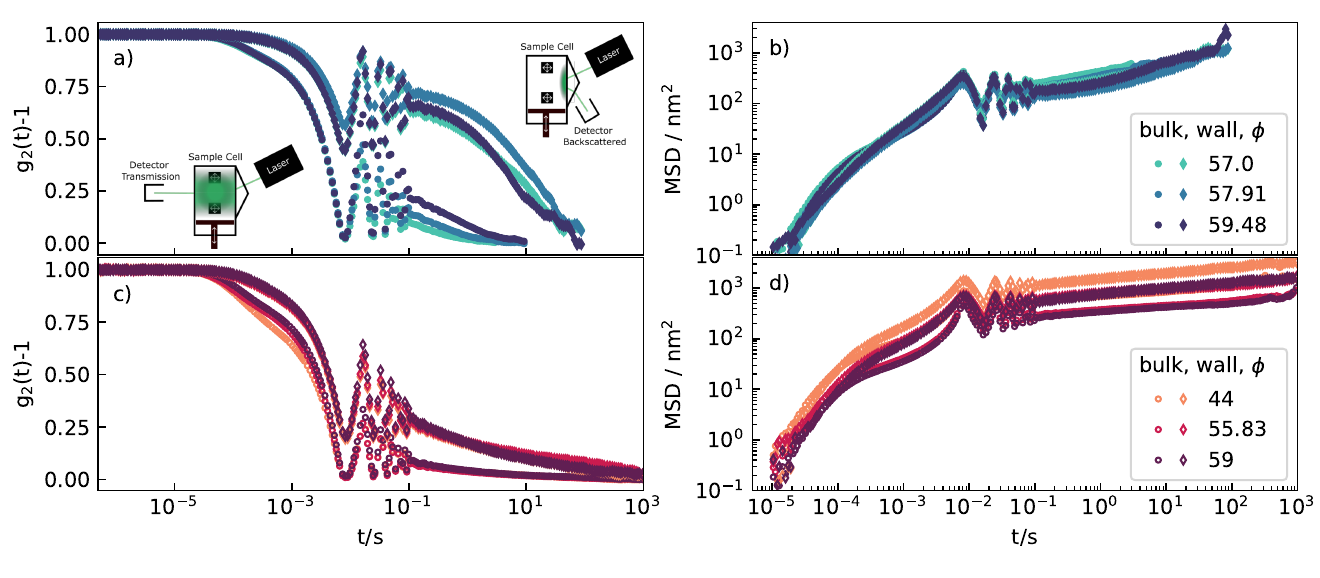}
    \caption{Intensity correlation functions and mean squared displacements for wall and bulk as indicated by the schematics in a) for ground measurements in a) and b) and for microgravity in c) and d).}
    \label{fig3}
\end{figure*}

We conducted similar experiments on the ISS and on ground with the same material, agitation frequency, and agitation time. Here in Fig.\ref{fig3}a and c we show the intensity correlation functions for ground and in microgravity as we gradually compress the sample. Notable features for all results are a decay of the correlation function with an oscillation imprinted to the decay. It can be seen that the first peak of the oscillation corresponds to the agitation frequency of approximately 60Hz. The results show correlation functions acquired over 10 hour measurement runs. This allows the full range of dynamics shown by the correlation function to be resolved. Shown are results in both the transmission and backscattering geometry. The volume fractions compared are different as on ground the minimal VF one can construct due to gravitational settling is 55\% where as in space our theoretical minimum based of our sample cell geometry was 44\%. In both cases, the bulk dynamics appears faster than the wall dynamic, e.g a faster decay. In the bulk dynamic on ground, a systematic slowing down is seen as the sample is compressed. This is less distinguishable for the microgravity sample, but a qualitative difference can be seen in the faster initial part of the decay. On ground, the wall results show a much slower decay than its counterpart on the space station. When viewing the ground results in the MSD representation, Fig.\ref{fig3}b, one can distinguish different regimes. A ballistic-like motion initially, followed by a plateau and finally sub-diffusive behavior shown by a power law exponent of 0.5. This is similar to results in microgravity, Fig.\ref{fig3}c, with ballistic behavior, this time without a recognizable plateau and a final sub-diffusive behavior that follows a power law exponent of 0.2. From the MSD representation on the ground, we also see that the measurements at the wall and in the bulk results tend to collapse to show the same curve. This is in contrast to the microgravity experiments, where the two dynamics show different behavior in the MSD representation.

\begin{figure}[th!]
    \centering
    \includegraphics[width=0.95\linewidth]{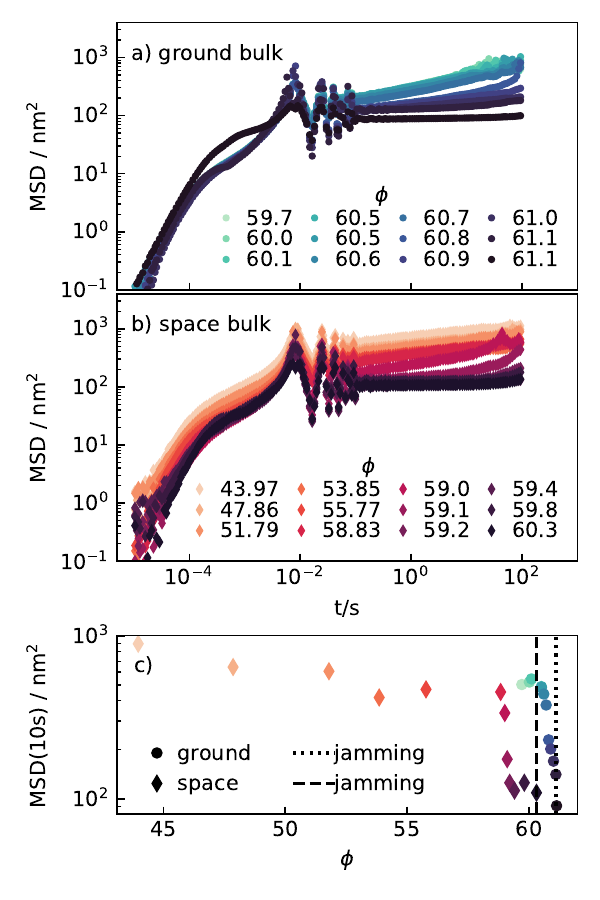}
    \caption{Volume fraction dependent mean squared displacement a) on the ground, b) in space and c) mean squared displacement at 10s depending on Volume fraction for ground and space illustrating glass transitions at 59.0\%$\pm$0.1 and 60.3\%$\pm$0.1 as well as jamming transition at 60.6\%$\pm$0.1 and 61.1\%$\pm$0.1.}
\label{fig4}
\end{figure}

With a basic understanding of the key differences of results in space and on ground, we can now extend our view over a larger range of VFs.  Fig.\ref{fig4} presents a case for identifying both the glass transition and then the resulting jamming for a set of experiment runs where the sample is systematically compressed and the intensity correlation functions are measured for 1 hour on the ISS and on ground. When analyzing Fig.\ref{fig4}a and \ref{fig4}b qualitatively, one can observe the slowing down of the dynamics as the sample gets denser for both ground and microgravity, shown by the decreasing MSD gradients. Fig.\ref{fig4}c. shows the MSDs at 10s, plotted as a function of the VF. The time of 10s is chosen due to it being the earliest position where the plateau appears clearly. At some VF, a significant decrease in MSD is observed. On ground this is at 60.6\%$\pm$0.1, and on the ISS at 59.0\%$\pm$0.1. The highest VF plotted for both sets of measurements represents the maximum densification point that was possible for that sample. With the influence of gravity, this point is 61.1\%$\pm$0.1, and in microgravity this is 60.3\%$\pm$0.1. Where the uncertainty is calculated from the uncertainty in the mass of the sample and the uncertainty in the encoder position.

\begin{figure}[th!]
\centering
\includegraphics[width=0.95\linewidth]{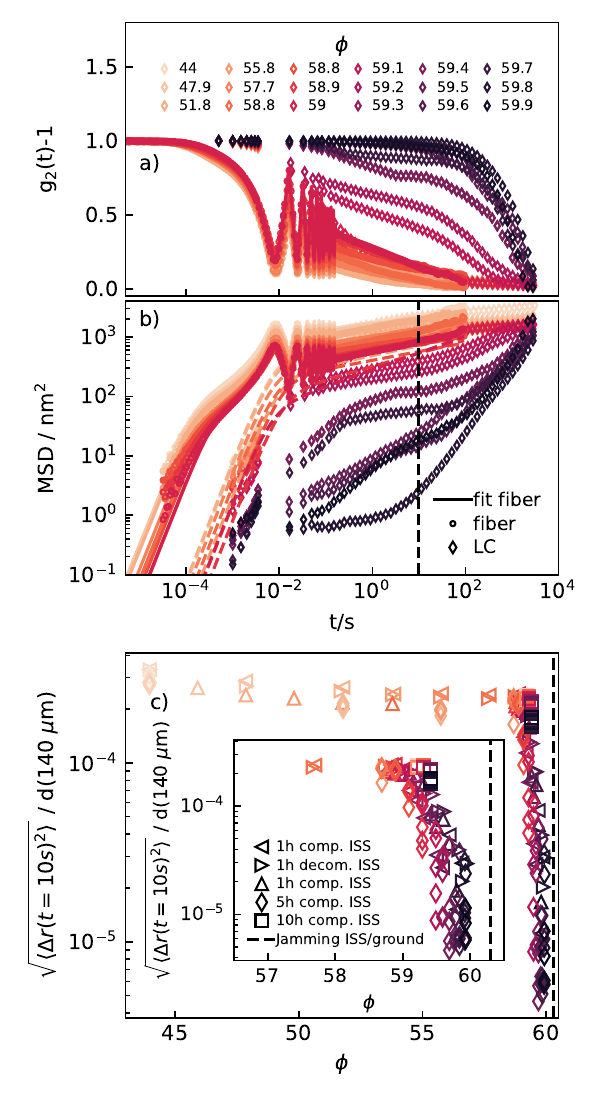}
\caption{Camera and fiber measurements in backscattering geometry presented as a) Intensity correlation functions and b) mean squared displacements, the dashed lines at 10s indicate the time where the plateau value is extracted for c) the volume fraction dependent comparison to root MSD normalized to the particle diameter.}
\label{fig5}
\end{figure}

The results in Fig.\ref{fig4} show only correlation functions formed from intensity fluctuations collected from the avalanche photodiodes (APDs) in the bulk dynamic. As mentioned, included in the experimental setup on the ISS is a line camera. This collects 500 individual speckles in the form of pixels and stitches them together in a time lapse. The resulting image shows a time evolution of change in speckle intensity with a time resolution of approximately 1ms. The intensities of these individual speckles are then correlated to form our correlation function. One of the key advantages of the line camera is the ability to easily remove the oscillations imprinted into the correlation function thereby showing the underlying particle dynamics. Fig.\ref{fig5} shows a full spectrum of VFs are measured starting from 45\% to a maximum of 61.3\%, from both the fiber and line camera. This figure shows a complete picture of our experimental campaign conducted on the ISS.  We again see the slowing of the long-time relaxation as the VF increases and the system becomes more dense. Fig.\ref{fig5}a and \ref{fig5}b includes all correlation functions and corresponding MSDs all of 1h, 5h, and 10 hour experiment duration runs in microgravity. When considering our fiber measurements, we see the dominating influence of the oscillation. But by fitting the data using the method outlined in the appendix, we can remove the oscillation which just leaves the underlying dynamic of the particles. This produces results from the fiber which match that of the line camera, lending credence to the reliability of our setup and the reproducibility of our results. Fig.\ref{fig5}c shows a calculated root mean square displacement as a fraction of the diameter of the particles, extracted from the line camera MSD. We can describe this as a localization length. These values are then plotted as a function of VF for all for expansion and densification runs. What is seen is what appears to be a state change in a range of VFs between 59\% and 59.5\%, which agree with the results seen in Fig.\ref{fig4}, where visually one can see that at that VF range the localization length starts to decrease significantly. This presents a clear change of state over a large set of experimental measurements in a microgravity environment. 

With the complete picture of the experimental results now presented, we can now discuss the interpretation of our results. First, let us address the difference in apparent wall and bulk dynamics.  It can be assumed that the light detected in backscattering geometry has been scattered a smaller number of times compared to in transmission\cite{furst2017microrheology}. The total number of scattering events make up the scattering path. With particle movement resulting in the decay of the correlation function. The shorter the length of the scattering path, as assumed when observing the wall dynamics, the less opportunities through particle movement are present to decay the correlation function. This then appears as a slower correlation function decay \cite{weitz1993diffusing}. This is seen in both microgravity and on ground. Our MSD representation takes into account our scattering path length difference described above, therefore negating the underlying reason for the difference in dynamics presented by the correlation functions. Our results in Fig.\ref{fig3}b on ground show that even with different looking correlation functions for the different scattering geometries, our results for the wall and bulk dynamics collapse into a single result. A difference remains in Fig.\ref{fig3}d  when looking at the MSDs for the wall and bulk dynamics in microgravity. From this one could infer that the system on ground is more homogeneous than in microgravity, as the dynamics shown by the MSD representation are similar at the wall and in the bulk. This could be due to it being easier to establish contacts between particles on ground and therefore transfer the energy produced by the piezos throughout the system more efficiently. Meaning particles at the boundaries of our sample cell experience a similar agitation to those in the majority of the bulk and therefore display similar dynamics\cite{andreotti2013granular}. 

Further analysis of the MSD representation shows that for both ground and space samples show sub-diffusive behavior at longer time scales with maximum MSD gradients of 0.5 and 0.2 respectively. This sub-diffusive behavior is similar to that seen in liquids\cite{Pabst2020,bohmer2024spectral}. Consequently, we can infer from our measurements that ground samples present behavior closer to a normal diffusion and a described Brownian motion. To reach this point we would either need a stronger agitation method as seen in \cite{Kunzner2025VC}, or to analyze longer length scales by probing with a laser with a longer wavelength, e.g. in the THz regime\cite{Born2014,Born2017}. But given our current setup we could infer that ground measurements are closer to reaching a state similar to a liquid on our measured time scale compared to in space. This is very intuitive as one would expect granular materiel to be more similar to a liquid without gravity because gravity is irrelevant for molecular liquids. Leaving an open question to be tackled.

Another open question concerns the length scale of particle dynamics that is probed in the experiment and the exact nature of these dynamics. We have shown that we have seemed to resolve the nanometer movement of 140-micrometer particles \cite{menon1997diffusing,menon1997particle}. We believe that the length scales calculated could describe two processes. First, we could measure the deformation of the particles on contact, as there is precedent in literature called clapping contacts. The deformations are on the order of 10 nm, depending on excitation energy, and stay in contact for about $1\cdot 10^{-6}$s \cite{rosato2020segregation,Hertz1881Contact,johnson1982one,mathey2024device}. This is not in our resolution limit but the calculation only considers normal contact forces, so this could still be responsible for the ballistic-like motion as shown in \cite{Kunzner2025VC}. Alternatively, this could be a true motion which is a combination of translation and rotation\cite{rosato2020segregation}. In which most of the sample exhibits minimal to no movement in comparison to the particles contacted directly by the piezos thereby reducing the average movement of the whole sample to nanometers\cite{mathey2024device}. Verification of these length scales comes in two-fold; first DWS theory tells us that our resolving length scale is proportional to the wavelength, and specifically the wavelength split across a lot of scattering events \cite{weitz1993diffusing,mie1908beitrage,Ni2013MSD,Xu1998MSD}. Secondly, other experiments that have looked at more dilute systems of similar-size particles also achieved length scales in this range \cite{furst2017microrheology,xing2018microrheology,cardinaux2002microrheology}. One of the key unknown factors in calculating our MSD is determining our l* or randomization length. This has been a challenge to verify experimentally with regard to light scattering on granular material. \cite{liu1998jamming,trappe2001jamming,xing2018microrheology,cardinaux2002microrheology,utermann2012friction}. There is also a disagreement when comparing length scales with those predicted from mode coupling theory (MCT) \cite{sperl2005nearly,sperl2012single}. But these calculations assume frictionless hard spheres, which presents a significant difference compared to our experiment. We hope some that future simulation work allows us to have a clearer picture regarding our length scales.

Now for the main objective of our investigation, determining both a glass and jamming transition for our samples. Within the field of granular matter, the processes of glass and jamming transitions vary in terms of their exact definitions \cite{charbonneau2017glass,coulais2013Jam,Behringer2015Jam,Silbert2002JamGlass}. We define a glass transition as the sudden change in the generalized dynamics of the granular system from a fluid, liquid-like state to that of an amorphous solid \cite{Coniglio2000Glass}. We define a jamming transition as an extreme extension into the realm of glassy dynamics, in which a mechanically stable state has been reached \cite{coulais2013Jam,goodrich2015scaling}, in our case above a timescale of 10s. Within the context of this experiment, we define a glass transition as when the dynamics start to dramatically decrease with increased density, and jamming is when the piston gets stuck and can no longer densify the system. For both of these transition points in microgravity, our results in Fig.\ref{fig5} show that the exact position is preparation dependent, as in the nature of the history of the densification process has an impact on the precise position of the transitions\cite{schroter2017local}. When similar agitation procedures are implemented both on ground and in microgravity, as shown in Fig.\ref{fig4}, two things become apparent. Firstly, the glass transition occurs at a higher density on ground, by 1.6\%. Which also provides the first experimental evidence of glassy dynamics in dense granular media in a microgravity environment. Secondly, the jamming transition point occurs 0.5\% later on ground than in space, which shows that on ground we can achieve a more dense system, as shown in previous experiments \cite{Olfa2022Microgravity}. Both for glass and jamming transitions positions still fall in the range of VFs predicted by \cite{Schröter_2007}.

So why does this occur? Well, in microgravity the cohesive inter-particle forces such as friction, Van der Waals and the electrostatic force \cite{Madel2021}, are the predominant forces acting on the system. These interactions limit the ability of the granular system to rearrange into more dense configurations compared to a system present in a gravitational force field, where gravity can apparently aid motion to overcame barriers with additional gravitational acceleration. This results in the earlier onset of glassy dynamics, due to the higher stability of each configuration in microgravity. With regards to jamming, the reduction in the re-arrangement ability better maintains the long force chains which permeate throughout the system to the boundaries and make the system mechanically stable \cite{Papa2007}, limiting the ability for the piston to compress the sample. 



In summary, we investigated a weak pulse-driven granular system under microgravity conditions on the ISS. We utilized DWS to describe the system by MSDs of the investigated polystyrene spheres. The periodically agitated particles show a sub-diffusive MSD behavior with a power law exponent of $\beta\approx0.5$ in contrast to the ISS data showing a slightly sub-diffusive behavior of $\beta\approx0.2$ which only changes to diffusive for a short time before the system is mechanically stuck or jammed. The sample cell creates volume fractions as low as 45\% on the ISS and has agitated the system continually for up to 10h. While the piston is densifying the sample, glassy dynamics are observed at volume fractions of $ \phi_{\textbf{g}}^{\textbf{iss}}>59.0\%$$\pm$0.1  compared to $\phi_{\textbf{g}}^{\textbf{gr}} >60.3\%$$\pm$0.1, which we believe is the first evidence of glassy dynamics experientially shown for a dense granular system in microgravity. Under the influence of gravity the jamming transition appears at a higher volume fraction, $\phi_{\textbf{jam}}^{\textbf{gr}} =61.1\%$$\pm$0.1, compared to in space, $\phi_{\textbf{jam}}^{\textbf{iss}}=60.6\%$$\pm$0.1. Showing we can achieve states on the ISS that are less dense but are as mechanically stable as denser states prepared on ground.   

\section{Acknowledgment}
This work was supported by the DLR Space Administration with funds provided by the Federal Ministry for Economic Affairs and Climate Action (BMWK) based on a decision of the German Federal Parliament under grant number 50WM1945 (SoMaDy2). We thank for the discussions with Philip Born, and Matthias Schröter.

\bibliography{ISS2024-bib}

\begin{thebibliography}{51}%
\makeatletter
\providecommand \@ifxundefined [1]{%
 \@ifx{#1\undefined}
}%
\providecommand \@ifnum [1]{%
 \ifnum #1\expandafter \@firstoftwo
 \else \expandafter \@secondoftwo
 \fi
}%
\providecommand \@ifx [1]{%
 \ifx #1\expandafter \@firstoftwo
 \else \expandafter \@secondoftwo
 \fi
}%
\providecommand \natexlab [1]{#1}%
\providecommand \enquote  [1]{``#1''}%
\providecommand \bibnamefont  [1]{#1}%
\providecommand \bibfnamefont [1]{#1}%
\providecommand \citenamefont [1]{#1}%
\providecommand \href@noop [0]{\@secondoftwo}%
\providecommand \href [0]{\begingroup \@sanitize@url \@href}%
\providecommand \@href[1]{\@@startlink{#1}\@@href}%
\providecommand \@@href[1]{\endgroup#1\@@endlink}%
\providecommand \@sanitize@url [0]{\catcode `\\12\catcode `\$12\catcode `\&12\catcode `\#12\catcode `\^12\catcode `\_12\catcode `\%12\relax}%
\providecommand \@@startlink[1]{}%
\providecommand \@@endlink[0]{}%
\providecommand \url  [0]{\begingroup\@sanitize@url \@url }%
\providecommand \@url [1]{\endgroup\@href {#1}{\urlprefix }}%
\providecommand \urlprefix  [0]{URL }%
\providecommand \Eprint [0]{\href }%
\providecommand \doibase [0]{https://doi.org/}%
\providecommand \selectlanguage [0]{\@gobble}%
\providecommand \bibinfo  [0]{\@secondoftwo}%
\providecommand \bibfield  [0]{\@secondoftwo}%
\providecommand \translation [1]{[#1]}%
\providecommand \BibitemOpen [0]{}%
\providecommand \bibitemStop [0]{}%
\providecommand \bibitemNoStop [0]{.\EOS\space}%
\providecommand \EOS [0]{\spacefactor3000\relax}%
\providecommand \BibitemShut  [1]{\csname bibitem#1\endcsname}%
\let\auto@bib@innerbib\@empty
\bibitem [{\citenamefont {Andreotti}\ \emph {et~al.}(2013)\citenamefont {Andreotti}, \citenamefont {Forterre},\ and\ \citenamefont {Pouliquen}}]{andreotti2013granular}%
  \BibitemOpen
  \bibfield  {author} {\bibinfo {author} {\bibfnamefont {B.}~\bibnamefont {Andreotti}}, \bibinfo {author} {\bibfnamefont {Y.}~\bibnamefont {Forterre}},\ and\ \bibinfo {author} {\bibfnamefont {O.}~\bibnamefont {Pouliquen}},\ }\href@noop {} {\emph {\bibinfo {title} {Granular media: between fluid and solid}}}\ (\bibinfo  {publisher} {Cambridge University Press},\ \bibinfo {year} {2013})\BibitemShut {NoStop}%
\bibitem [{\citenamefont {Duran}(2012)}]{duran2012sands}%
  \BibitemOpen
  \bibfield  {author} {\bibinfo {author} {\bibfnamefont {J.}~\bibnamefont {Duran}},\ }\href@noop {} {\emph {\bibinfo {title} {Sands, powders, and grains: an introduction to the physics of granular materials}}}\ (\bibinfo  {publisher} {Springer Science \& Business Media},\ \bibinfo {year} {2012})\BibitemShut {NoStop}%
\bibitem [{\citenamefont {Duran}(1997)}]{duran1997sables}%
  \BibitemOpen
  \bibfield  {author} {\bibinfo {author} {\bibfnamefont {J.}~\bibnamefont {Duran}},\ }\href@noop {} {\emph {\bibinfo {title} {Sables, poudres et grains}}},\ \bibinfo {number} {BOOK}\ (\bibinfo  {publisher} {Eyrolles},\ \bibinfo {year} {1997})\BibitemShut {NoStop}%
\bibitem [{\citenamefont {Janssen}(1895)}]{janssen1895versuche}%
  \BibitemOpen
  \bibfield  {author} {\bibinfo {author} {\bibfnamefont {H.}~\bibnamefont {Janssen}},\ }\bibfield  {title} {\bibinfo {title} {Versuche uber getreidedruck in silozellen},\ }\href@noop {} {\bibfield  {journal} {\bibinfo  {journal} {Z. ver. deut. Ing.}\ }\textbf {\bibinfo {volume} {39}},\ \bibinfo {pages} {1045} (\bibinfo {year} {1895})}\BibitemShut {NoStop}%
\bibitem [{\citenamefont {Windows-Yule}\ \emph {et~al.}(2019)\citenamefont {Windows-Yule}, \citenamefont {M{\"u}hlbauer}, \citenamefont {Cisneros}, \citenamefont {Nair}, \citenamefont {Marzulli},\ and\ \citenamefont {P{\"o}schel}}]{windows2019janssen}%
  \BibitemOpen
  \bibfield  {author} {\bibinfo {author} {\bibfnamefont {C.}~\bibnamefont {Windows-Yule}}, \bibinfo {author} {\bibfnamefont {S.}~\bibnamefont {M{\"u}hlbauer}}, \bibinfo {author} {\bibfnamefont {L.~T.}\ \bibnamefont {Cisneros}}, \bibinfo {author} {\bibfnamefont {P.}~\bibnamefont {Nair}}, \bibinfo {author} {\bibfnamefont {V.}~\bibnamefont {Marzulli}},\ and\ \bibinfo {author} {\bibfnamefont {T.}~\bibnamefont {P{\"o}schel}},\ }\bibfield  {title} {\bibinfo {title} {Janssen effect in dynamic particulate systems},\ }\href@noop {} {\bibfield  {journal} {\bibinfo  {journal} {Physical Review E}\ }\textbf {\bibinfo {volume} {100}},\ \bibinfo {pages} {022902} (\bibinfo {year} {2019})}\BibitemShut {NoStop}%
\bibitem [{\citenamefont {Wentworth}(1922)}]{wentworth1922scale}%
  \BibitemOpen
  \bibfield  {author} {\bibinfo {author} {\bibfnamefont {C.~K.}\ \bibnamefont {Wentworth}},\ }\bibfield  {title} {\bibinfo {title} {A scale of grade and class terms for clastic sediments},\ }\href@noop {} {\bibfield  {journal} {\bibinfo  {journal} {The journal of geology}\ }\textbf {\bibinfo {volume} {30}},\ \bibinfo {pages} {377} (\bibinfo {year} {1922})}\BibitemShut {NoStop}%
\bibitem [{\citenamefont {Rosato}\ and\ \citenamefont {Windows-Yule}(2020)}]{rosato2020segregation}%
  \BibitemOpen
  \bibfield  {author} {\bibinfo {author} {\bibfnamefont {A.~D.}\ \bibnamefont {Rosato}}\ and\ \bibinfo {author} {\bibfnamefont {C.}~\bibnamefont {Windows-Yule}},\ }\href@noop {} {\emph {\bibinfo {title} {Segregation in vibrated granular systems}}}\ (\bibinfo  {publisher} {Academic Press},\ \bibinfo {year} {2020})\BibitemShut {NoStop}%
\bibitem [{\citenamefont {Opsomer}\ \emph {et~al.}(2011)\citenamefont {Opsomer}, \citenamefont {Ludewig},\ and\ \citenamefont {Vandewalle}}]{Opsommer2011}%
  \BibitemOpen
  \bibfield  {author} {\bibinfo {author} {\bibfnamefont {E.}~\bibnamefont {Opsomer}}, \bibinfo {author} {\bibfnamefont {F.}~\bibnamefont {Ludewig}},\ and\ \bibinfo {author} {\bibfnamefont {N.}~\bibnamefont {Vandewalle}},\ }\bibfield  {title} {\bibinfo {title} {Phase transitions in vibrated granular systems in microgravity},\ }\href {https://doi.org/10.1103/PhysRevE.84.051306} {\bibfield  {journal} {\bibinfo  {journal} {Phys. Rev. E}\ }\textbf {\bibinfo {volume} {84}},\ \bibinfo {pages} {051306} (\bibinfo {year} {2011})}\BibitemShut {NoStop}%
\bibitem [{\citenamefont {Hou}\ \emph {et~al.}(2008)\citenamefont {Hou}, \citenamefont {Liu}, \citenamefont {Zhai}, \citenamefont {Sun}, \citenamefont {Lu}, \citenamefont {Garrabos},\ and\ \citenamefont {Evesque}}]{Evesque}%
  \BibitemOpen
  \bibfield  {author} {\bibinfo {author} {\bibfnamefont {M.}~\bibnamefont {Hou}}, \bibinfo {author} {\bibfnamefont {R.}~\bibnamefont {Liu}}, \bibinfo {author} {\bibfnamefont {G.}~\bibnamefont {Zhai}}, \bibinfo {author} {\bibfnamefont {Z.}~\bibnamefont {Sun}}, \bibinfo {author} {\bibfnamefont {K.}~\bibnamefont {Lu}}, \bibinfo {author} {\bibfnamefont {Y.}~\bibnamefont {Garrabos}},\ and\ \bibinfo {author} {\bibfnamefont {P.}~\bibnamefont {Evesque}},\ }\bibfield  {title} {\bibinfo {title} {Velocity distribution of vibration-driven granular gas in knudsen regime in microgravity},\ }\href {https://doi.org/10.1007/s12217-008-9040-5} {\bibfield  {journal} {\bibinfo  {journal} {Microgravity Science and Technology}\ }\textbf {\bibinfo {volume} {20}},\ \bibinfo {pages} {73} (\bibinfo {year} {2008})}\BibitemShut {NoStop}%
\bibitem [{\citenamefont {Wang}\ \emph {et~al.}(2018)\citenamefont {Wang}, \citenamefont {Hou}, \citenamefont {Chen}, \citenamefont {Yu},\ and\ \citenamefont {Sperl}}]{Wang}%
  \BibitemOpen
  \bibfield  {author} {\bibinfo {author} {\bibfnamefont {W.-G.}\ \bibnamefont {Wang}}, \bibinfo {author} {\bibfnamefont {M.-Y.}\ \bibnamefont {Hou}}, \bibinfo {author} {\bibfnamefont {K.}~\bibnamefont {Chen}}, \bibinfo {author} {\bibfnamefont {P.-D.}\ \bibnamefont {Yu}},\ and\ \bibinfo {author} {\bibfnamefont {M.}~\bibnamefont {Sperl}},\ }\bibfield  {title} {\bibinfo {title} {Experimental and numerical study on energy dissipation in freely cooling granular gases under microgravity},\ }\href@noop {} {\bibfield  {journal} {\bibinfo  {journal} {Chinese Physics B}\ }\textbf {\bibinfo {volume} {27}},\ \bibinfo {pages} {084501} (\bibinfo {year} {2018})}\BibitemShut {NoStop}%
\bibitem [{\citenamefont {Pitikaris}\ \emph {et~al.}(2022)\citenamefont {Pitikaris}, \citenamefont {Bartz}, \citenamefont {Yu}, \citenamefont {Cristoforetti},\ and\ \citenamefont {Sperl}}]{Piti2022}%
  \BibitemOpen
  \bibfield  {author} {\bibinfo {author} {\bibfnamefont {S.}~\bibnamefont {Pitikaris}}, \bibinfo {author} {\bibfnamefont {P.}~\bibnamefont {Bartz}}, \bibinfo {author} {\bibfnamefont {P.}~\bibnamefont {Yu}}, \bibinfo {author} {\bibfnamefont {S.}~\bibnamefont {Cristoforetti}},\ and\ \bibinfo {author} {\bibfnamefont {M.}~\bibnamefont {Sperl}},\ }\bibfield  {title} {\bibinfo {title} {Granular cooling of ellipsoidal particles in microgravity},\ }\href {https://doi.org/10.1038/s41526-022-00196-6} {\bibfield  {journal} {\bibinfo  {journal} {npj Microgravity}\ }\textbf {\bibinfo {volume} {8}},\ \bibinfo {pages} {11} (\bibinfo {year} {2022})}\BibitemShut {NoStop}%
\bibitem [{\citenamefont {Born}\ \emph {et~al.}(2021)\citenamefont {Born}, \citenamefont {Braibanti}, \citenamefont {Cristofolini}, \citenamefont {Cohen-Addad}, \citenamefont {Durian}, \citenamefont {Egelhaaf}, \citenamefont {Escobedo-S{\'a}nchez}, \citenamefont {H{\"o}hler}, \citenamefont {Karapantsios}, \citenamefont {Langevin} \emph {et~al.}}]{born2021soft}%
  \BibitemOpen
  \bibfield  {author} {\bibinfo {author} {\bibfnamefont {P.}~\bibnamefont {Born}}, \bibinfo {author} {\bibfnamefont {M.}~\bibnamefont {Braibanti}}, \bibinfo {author} {\bibfnamefont {L.}~\bibnamefont {Cristofolini}}, \bibinfo {author} {\bibfnamefont {S.}~\bibnamefont {Cohen-Addad}}, \bibinfo {author} {\bibfnamefont {D.}~\bibnamefont {Durian}}, \bibinfo {author} {\bibfnamefont {S.}~\bibnamefont {Egelhaaf}}, \bibinfo {author} {\bibfnamefont {M.}~\bibnamefont {Escobedo-S{\'a}nchez}}, \bibinfo {author} {\bibfnamefont {R.}~\bibnamefont {H{\"o}hler}}, \bibinfo {author} {\bibfnamefont {T.}~\bibnamefont {Karapantsios}}, \bibinfo {author} {\bibfnamefont {D.}~\bibnamefont {Langevin}}, \emph {et~al.},\ }\bibfield  {title} {\bibinfo {title} {Soft matter dynamics: A versatile microgravity platform to study dynamics in soft matter},\ }\href@noop {} {\bibfield  {journal} {\bibinfo  {journal} {Review of Scientific Instruments}\ }\textbf {\bibinfo {volume} {92}} (\bibinfo {year} {2021})}\BibitemShut {NoStop}%
\bibitem [{\citenamefont {Weitz}\ \emph {et~al.}(1993)\citenamefont {Weitz}, \citenamefont {Zhu}, \citenamefont {Durian}, \citenamefont {Gang},\ and\ \citenamefont {Pine}}]{weitz1993diffusing}%
  \BibitemOpen
  \bibfield  {author} {\bibinfo {author} {\bibfnamefont {D.}~\bibnamefont {Weitz}}, \bibinfo {author} {\bibfnamefont {J.}~\bibnamefont {Zhu}}, \bibinfo {author} {\bibfnamefont {D.}~\bibnamefont {Durian}}, \bibinfo {author} {\bibfnamefont {H.}~\bibnamefont {Gang}},\ and\ \bibinfo {author} {\bibfnamefont {D.}~\bibnamefont {Pine}},\ }\bibfield  {title} {\bibinfo {title} {Diffusing-wave spectroscopy: The technique and some applications},\ }\href@noop {} {\bibfield  {journal} {\bibinfo  {journal} {Physica Scripta}\ }\textbf {\bibinfo {volume} {1993}},\ \bibinfo {pages} {610} (\bibinfo {year} {1993})}\BibitemShut {NoStop}%
\bibitem [{\citenamefont {Brown}(1993)}]{Brown1993Dynamic}%
  \BibitemOpen
  \bibfield  {author} {\bibinfo {author} {\bibfnamefont {W.}~\bibnamefont {Brown}},\ }\bibinfo {title} {Dynamic light scattering}\ (\bibinfo  {publisher} {Clarendon Press},\ \bibinfo {year} {1993})\ Chap.~\bibinfo {chapter} {16}\BibitemShut {NoStop}%
\bibitem [{\citenamefont {Kunzner}(2025)}]{Kunzner2025VC}%
  \BibitemOpen
  \bibfield  {author} {\bibinfo {author} {\bibfnamefont {M.}~\bibnamefont {Kunzner}},\ }\bibfield  {title} {\bibinfo {title} {The dynamics in vibro-fluidized beds: A diffusing wave spectroscopy study},\ }\href@noop {} {\bibfield  {journal} {\bibinfo  {journal} {arXiv: https://doi.org/10.48550/arXiv.2503.00517}\ } (\bibinfo {year} {2025})}\BibitemShut {NoStop}%
\bibitem [{\citenamefont {Trappe}\ \emph {et~al.}(2001)\citenamefont {Trappe}, \citenamefont {Prasad}, \citenamefont {Cipelletti}, \citenamefont {Segre},\ and\ \citenamefont {Weitz}}]{trappe2001jamming}%
  \BibitemOpen
  \bibfield  {author} {\bibinfo {author} {\bibfnamefont {V.}~\bibnamefont {Trappe}}, \bibinfo {author} {\bibfnamefont {V.}~\bibnamefont {Prasad}}, \bibinfo {author} {\bibfnamefont {L.}~\bibnamefont {Cipelletti}}, \bibinfo {author} {\bibfnamefont {P.}~\bibnamefont {Segre}},\ and\ \bibinfo {author} {\bibfnamefont {D.~A.}\ \bibnamefont {Weitz}},\ }\bibfield  {title} {\bibinfo {title} {Jamming phase diagram for attractive particles},\ }\href@noop {} {\bibfield  {journal} {\bibinfo  {journal} {Nature}\ }\textbf {\bibinfo {volume} {411}},\ \bibinfo {pages} {772} (\bibinfo {year} {2001})}\BibitemShut {NoStop}%
\bibitem [{\citenamefont {Liu}\ and\ \citenamefont {Nagel}(1998)}]{liu1998jamming}%
  \BibitemOpen
  \bibfield  {author} {\bibinfo {author} {\bibfnamefont {A.~J.}\ \bibnamefont {Liu}}\ and\ \bibinfo {author} {\bibfnamefont {S.~R.}\ \bibnamefont {Nagel}},\ }\bibfield  {title} {\bibinfo {title} {Jamming is not just cool any more},\ }\href@noop {} {\bibfield  {journal} {\bibinfo  {journal} {Nature}\ }\textbf {\bibinfo {volume} {396}},\ \bibinfo {pages} {21} (\bibinfo {year} {1998})}\BibitemShut {NoStop}%
\bibitem [{\citenamefont {Berthier}\ and\ \citenamefont {Biroli}(2009)}]{berthier2009glasses}%
  \BibitemOpen
  \bibfield  {author} {\bibinfo {author} {\bibfnamefont {L.}~\bibnamefont {Berthier}}\ and\ \bibinfo {author} {\bibfnamefont {G.}~\bibnamefont {Biroli}},\ }\href@noop {} {\bibinfo {title} {Glasses and aging, a statistical mechanics perspective on.}} (\bibinfo {year} {2009})\BibitemShut {NoStop}%
\bibitem [{\citenamefont {Happel}(1993)}]{happel1993indigenous}%
  \BibitemOpen
  \bibfield  {author} {\bibinfo {author} {\bibfnamefont {J.~A.}\ \bibnamefont {Happel}},\ }\bibfield  {title} {\bibinfo {title} {Indigenous materials for lunar construction},\ }\href@noop {} {\bibfield  {journal} {\bibinfo  {journal} {Applied Mechanics Reviews}\ } (\bibinfo {year} {1993})}\BibitemShut {NoStop}%
\bibitem [{\citenamefont {Cumberland}\ and\ \citenamefont {Crawford}(1987)}]{cumberland1987packing}%
  \BibitemOpen
  \bibfield  {author} {\bibinfo {author} {\bibfnamefont {D.}~\bibnamefont {Cumberland}}\ and\ \bibinfo {author} {\bibfnamefont {R.~J.}\ \bibnamefont {Crawford}},\ }\href@noop {} {\emph {\bibinfo {title} {The packing of particles}}}\ (\bibinfo  {publisher} {Elsevier Science Pub. Co. Inc., New York, NY},\ \bibinfo {year} {1987})\BibitemShut {NoStop}%
\bibitem [{\citenamefont {Thi\'evenaz}\ and\ \citenamefont {Sauret}(2021)}]{PhysRevFluids.6.L062301}%
  \BibitemOpen
  \bibfield  {author} {\bibinfo {author} {\bibfnamefont {V.}~\bibnamefont {Thi\'evenaz}}\ and\ \bibinfo {author} {\bibfnamefont {A.}~\bibnamefont {Sauret}},\ }\bibfield  {title} {\bibinfo {title} {Pinch-off of viscoelastic particulate suspensions},\ }\href {https://doi.org/10.1103/PhysRevFluids.6.L062301} {\bibfield  {journal} {\bibinfo  {journal} {Phys. Rev. Fluids}\ }\textbf {\bibinfo {volume} {6}},\ \bibinfo {pages} {L062301} (\bibinfo {year} {2021})}\BibitemShut {NoStop}%
\bibitem [{\citenamefont {Berne}\ and\ \citenamefont {Pecora}(2000)}]{berne2000dynamic}%
  \BibitemOpen
  \bibfield  {author} {\bibinfo {author} {\bibfnamefont {B.~J.}\ \bibnamefont {Berne}}\ and\ \bibinfo {author} {\bibfnamefont {R.}~\bibnamefont {Pecora}},\ }\href@noop {} {\emph {\bibinfo {title} {Dynamic light scattering: with applications to chemistry, biology, and physics}}}\ (\bibinfo  {publisher} {Courier Corporation},\ \bibinfo {year} {2000})\BibitemShut {NoStop}%
\bibitem [{\citenamefont {Furst}\ and\ \citenamefont {Squires}(2017)}]{furst2017microrheology}%
  \BibitemOpen
  \bibfield  {author} {\bibinfo {author} {\bibfnamefont {E.~M.}\ \bibnamefont {Furst}}\ and\ \bibinfo {author} {\bibfnamefont {T.~M.}\ \bibnamefont {Squires}},\ }\href@noop {} {\emph {\bibinfo {title} {Microrheology}}}\ (\bibinfo  {publisher} {Oxford University Press},\ \bibinfo {year} {2017})\BibitemShut {NoStop}%
\bibitem [{\citenamefont {Pabst}\ \emph {et~al.}(2020)\citenamefont {Pabst}, \citenamefont {Helbling}, \citenamefont {Gabriel}, \citenamefont {Weigl},\ and\ \citenamefont {Blochowicz}}]{Pabst2020}%
  \BibitemOpen
  \bibfield  {author} {\bibinfo {author} {\bibfnamefont {F.}~\bibnamefont {Pabst}}, \bibinfo {author} {\bibfnamefont {A.}~\bibnamefont {Helbling}}, \bibinfo {author} {\bibfnamefont {J.}~\bibnamefont {Gabriel}}, \bibinfo {author} {\bibfnamefont {P.}~\bibnamefont {Weigl}},\ and\ \bibinfo {author} {\bibfnamefont {T.}~\bibnamefont {Blochowicz}},\ }\bibfield  {title} {\bibinfo {title} {Dipole-dipole correlations and the debye process in the dielectric response of nonassociating glass forming liquids},\ }\href {https://doi.org/10.1103/PhysRevE.102.010606} {\bibfield  {journal} {\bibinfo  {journal} {Phys. Rev. E}\ }\textbf {\bibinfo {volume} {102}},\ \bibinfo {pages} {010606} (\bibinfo {year} {2020})}\BibitemShut {NoStop}%
\bibitem [{\citenamefont {B{\"o}hmer}\ \emph {et~al.}(2024)\citenamefont {B{\"o}hmer}, \citenamefont {Pabst}, \citenamefont {Gabriel}, \citenamefont {Zei{\ss}ler},\ and\ \citenamefont {Blochowicz}}]{bohmer2024spectral}%
  \BibitemOpen
  \bibfield  {author} {\bibinfo {author} {\bibfnamefont {T.}~\bibnamefont {B{\"o}hmer}}, \bibinfo {author} {\bibfnamefont {F.}~\bibnamefont {Pabst}}, \bibinfo {author} {\bibfnamefont {J.}~\bibnamefont {Gabriel}}, \bibinfo {author} {\bibfnamefont {R.}~\bibnamefont {Zei{\ss}ler}},\ and\ \bibinfo {author} {\bibfnamefont {T.}~\bibnamefont {Blochowicz}},\ }\bibfield  {title} {\bibinfo {title} {On the spectral shape of the structural relaxation in deeply supercooled liquids},\ }\href@noop {} {\bibfield  {journal} {\bibinfo  {journal} {arXiv preprint arXiv:2412.17014}\ } (\bibinfo {year} {2024})}\BibitemShut {NoStop}%
\bibitem [{\citenamefont {Born}\ \emph {et~al.}(2014)\citenamefont {Born}, \citenamefont {Rothbart}, \citenamefont {Sperl},\ and\ \citenamefont {Hübers}}]{Born2014}%
  \BibitemOpen
  \bibfield  {author} {\bibinfo {author} {\bibfnamefont {P.}~\bibnamefont {Born}}, \bibinfo {author} {\bibfnamefont {N.}~\bibnamefont {Rothbart}}, \bibinfo {author} {\bibfnamefont {M.}~\bibnamefont {Sperl}},\ and\ \bibinfo {author} {\bibfnamefont {H.-W.}\ \bibnamefont {Hübers}},\ }\bibfield  {title} {\bibinfo {title} {Granular structure determined by terahertz scattering},\ }\href {https://doi.org/10.1209/0295-5075/106/48006} {\bibfield  {journal} {\bibinfo  {journal} {Europhysics Letters}\ }\textbf {\bibinfo {volume} {106}},\ \bibinfo {pages} {48006} (\bibinfo {year} {2014})}\BibitemShut {NoStop}%
\bibitem [{\citenamefont {Born}\ and\ \citenamefont {Holldack}(2017)}]{Born2017}%
  \BibitemOpen
  \bibfield  {author} {\bibinfo {author} {\bibfnamefont {P.}~\bibnamefont {Born}}\ and\ \bibinfo {author} {\bibfnamefont {K.}~\bibnamefont {Holldack}},\ }\bibfield  {title} {\bibinfo {title} {Analysis of granular packing structure by scattering of thz radiation},\ }\href {https://doi.org/10.1063/1.4983045} {\bibfield  {journal} {\bibinfo  {journal} {Review of Scientific Instruments}\ }\textbf {\bibinfo {volume} {88}},\ \bibinfo {pages} {051802} (\bibinfo {year} {2017})}\BibitemShut {NoStop}%
\bibitem [{\citenamefont {Menon}\ and\ \citenamefont {Durian}(1997{\natexlab{a}})}]{menon1997diffusing}%
  \BibitemOpen
  \bibfield  {author} {\bibinfo {author} {\bibfnamefont {N.}~\bibnamefont {Menon}}\ and\ \bibinfo {author} {\bibfnamefont {D.~J.}\ \bibnamefont {Durian}},\ }\bibfield  {title} {\bibinfo {title} {Diffusing-wave spectroscopy of dynamics in a three-dimensional granular flow},\ }\href@noop {} {\bibfield  {journal} {\bibinfo  {journal} {Science}\ }\textbf {\bibinfo {volume} {275}},\ \bibinfo {pages} {1920} (\bibinfo {year} {1997}{\natexlab{a}})}\BibitemShut {NoStop}%
\bibitem [{\citenamefont {Menon}\ and\ \citenamefont {Durian}(1997{\natexlab{b}})}]{menon1997particle}%
  \BibitemOpen
  \bibfield  {author} {\bibinfo {author} {\bibfnamefont {N.}~\bibnamefont {Menon}}\ and\ \bibinfo {author} {\bibfnamefont {D.~J.}\ \bibnamefont {Durian}},\ }\bibfield  {title} {\bibinfo {title} {Particle motions in a gas-fluidized bed of sand},\ }\href@noop {} {\bibfield  {journal} {\bibinfo  {journal} {Physical Review Letters}\ }\textbf {\bibinfo {volume} {79}},\ \bibinfo {pages} {3407} (\bibinfo {year} {1997}{\natexlab{b}})}\BibitemShut {NoStop}%
\bibitem [{\citenamefont {Hertz}(1881)}]{Hertz1881Contact}%
  \BibitemOpen
  \bibfield  {author} {\bibinfo {author} {\bibfnamefont {H.}~\bibnamefont {Hertz}},\ }\bibfield  {title} {\bibinfo {title} {Über die berührung fester elastischer körper},\ }\href@noop {} {\bibfield  {journal} {\bibinfo  {journal} {Journal für die reine und angewandte Mathematik}\ }\textbf {\bibinfo {volume} {92}},\ \bibinfo {pages} {156} (\bibinfo {year} {1881})}\BibitemShut {NoStop}%
\bibitem [{\citenamefont {Johnson}(1982)}]{johnson1982one}%
  \BibitemOpen
  \bibfield  {author} {\bibinfo {author} {\bibfnamefont {K.~L.}\ \bibnamefont {Johnson}},\ }\bibfield  {title} {\bibinfo {title} {One hundred years of hertz contact},\ }\href@noop {} {\bibfield  {journal} {\bibinfo  {journal} {Proceedings of the Institution of Mechanical Engineers}\ }\textbf {\bibinfo {volume} {196}},\ \bibinfo {pages} {363} (\bibinfo {year} {1982})}\BibitemShut {NoStop}%
\bibitem [{\citenamefont {Mathey}\ \emph {et~al.}(2024)\citenamefont {Mathey}, \citenamefont {Fur}, \citenamefont {Chasle}, \citenamefont {Amon},\ and\ \citenamefont {Crassous}}]{mathey2024device}%
  \BibitemOpen
  \bibfield  {author} {\bibinfo {author} {\bibfnamefont {A.}~\bibnamefont {Mathey}}, \bibinfo {author} {\bibfnamefont {M.~L.}\ \bibnamefont {Fur}}, \bibinfo {author} {\bibfnamefont {P.}~\bibnamefont {Chasle}}, \bibinfo {author} {\bibfnamefont {A.}~\bibnamefont {Amon}},\ and\ \bibinfo {author} {\bibfnamefont {J.}~\bibnamefont {Crassous}},\ }\bibfield  {title} {\bibinfo {title} {A device for studying elementary plasticity fluctuations in granular media},\ }\href@noop {} {\bibfield  {journal} {\bibinfo  {journal} {arXiv preprint arXiv:2403.09396}\ } (\bibinfo {year} {2024})}\BibitemShut {NoStop}%
\bibitem [{\citenamefont {Mie}(1908)}]{mie1908beitrage}%
  \BibitemOpen
  \bibfield  {author} {\bibinfo {author} {\bibfnamefont {G.}~\bibnamefont {Mie}},\ }\bibfield  {title} {\bibinfo {title} {Beitr{\"a}ge zur optik tr{\"u}ber medien, speziell kolloidaler metall{\"o}sungen},\ }\href@noop {} {\bibfield  {journal} {\bibinfo  {journal} {Annalen der physik}\ }\textbf {\bibinfo {volume} {330}},\ \bibinfo {pages} {377} (\bibinfo {year} {1908})}\BibitemShut {NoStop}%
\bibitem [{\citenamefont {Ni}\ \emph {et~al.}(2013)\citenamefont {Ni}, \citenamefont {Cohen~Stuart},\ and\ \citenamefont {Dijkstra}}]{Ni2013MSD}%
  \BibitemOpen
  \bibfield  {author} {\bibinfo {author} {\bibfnamefont {R.}~\bibnamefont {Ni}}, \bibinfo {author} {\bibfnamefont {M.}~\bibnamefont {Cohen~Stuart}},\ and\ \bibinfo {author} {\bibfnamefont {M.}~\bibnamefont {Dijkstra}},\ }\bibfield  {title} {\bibinfo {title} {Pushing the glass transition towards random close packing using self-propelled hard spheres},\ }\href {https://doi.org/10.1038/ncomms3704} {\bibfield  {journal} {\bibinfo  {journal} {Nature communications}\ }\textbf {\bibinfo {volume} {4}},\ \bibinfo {pages} {2704} (\bibinfo {year} {2013})}\BibitemShut {NoStop}%
\bibitem [{\citenamefont {Xu}\ \emph {et~al.}(1998)\citenamefont {Xu}, \citenamefont {Viasnoff},\ and\ \citenamefont {Wirtz}}]{Xu1998MSD}%
  \BibitemOpen
  \bibfield  {author} {\bibinfo {author} {\bibfnamefont {J.}~\bibnamefont {Xu}}, \bibinfo {author} {\bibfnamefont {V.}~\bibnamefont {Viasnoff}},\ and\ \bibinfo {author} {\bibfnamefont {D.}~\bibnamefont {Wirtz}},\ }\bibfield  {title} {\bibinfo {title} {Compliance of actin filament networks measured by particle-tracking microrheology and diffusing wave spectroscopy},\ }\href {https://doi.org/10.1007/s003970050125} {\bibfield  {journal} {\bibinfo  {journal} {Rheologica Acta}\ }\textbf {\bibinfo {volume} {37}},\ \bibinfo {pages} {387} (\bibinfo {year} {1998})}\BibitemShut {NoStop}%
\bibitem [{\citenamefont {Xing}\ \emph {et~al.}(2018)\citenamefont {Xing}, \citenamefont {Caciagli}, \citenamefont {Cao}, \citenamefont {Stoev}, \citenamefont {Zupkauskas}, \citenamefont {O’Neill}, \citenamefont {Wenzel}, \citenamefont {Lamboll}, \citenamefont {Liu},\ and\ \citenamefont {Eiser}}]{xing2018microrheology}%
  \BibitemOpen
  \bibfield  {author} {\bibinfo {author} {\bibfnamefont {Z.}~\bibnamefont {Xing}}, \bibinfo {author} {\bibfnamefont {A.}~\bibnamefont {Caciagli}}, \bibinfo {author} {\bibfnamefont {T.}~\bibnamefont {Cao}}, \bibinfo {author} {\bibfnamefont {I.}~\bibnamefont {Stoev}}, \bibinfo {author} {\bibfnamefont {M.}~\bibnamefont {Zupkauskas}}, \bibinfo {author} {\bibfnamefont {T.}~\bibnamefont {O’Neill}}, \bibinfo {author} {\bibfnamefont {T.}~\bibnamefont {Wenzel}}, \bibinfo {author} {\bibfnamefont {R.}~\bibnamefont {Lamboll}}, \bibinfo {author} {\bibfnamefont {D.}~\bibnamefont {Liu}},\ and\ \bibinfo {author} {\bibfnamefont {E.}~\bibnamefont {Eiser}},\ }\bibfield  {title} {\bibinfo {title} {Microrheology of dna hydrogels},\ }\href@noop {} {\bibfield  {journal} {\bibinfo  {journal} {Proceedings of the National Academy of Sciences}\ }\textbf {\bibinfo {volume} {115}},\ \bibinfo {pages} {8137} (\bibinfo {year} {2018})}\BibitemShut {NoStop}%
\bibitem [{\citenamefont {Cardinaux}\ \emph {et~al.}(2002)\citenamefont {Cardinaux}, \citenamefont {Cipelletti}, \citenamefont {Scheffold},\ and\ \citenamefont {Schurtenberger}}]{cardinaux2002microrheology}%
  \BibitemOpen
  \bibfield  {author} {\bibinfo {author} {\bibfnamefont {F.}~\bibnamefont {Cardinaux}}, \bibinfo {author} {\bibfnamefont {L.}~\bibnamefont {Cipelletti}}, \bibinfo {author} {\bibfnamefont {F.}~\bibnamefont {Scheffold}},\ and\ \bibinfo {author} {\bibfnamefont {P.}~\bibnamefont {Schurtenberger}},\ }\bibfield  {title} {\bibinfo {title} {Microrheology of giant-micelle solutions},\ }\href@noop {} {\bibfield  {journal} {\bibinfo  {journal} {Europhysics Letters}\ }\textbf {\bibinfo {volume} {57}},\ \bibinfo {pages} {738} (\bibinfo {year} {2002})}\BibitemShut {NoStop}%
\bibitem [{\citenamefont {Utermann}(2012)}]{utermann2012friction}%
  \BibitemOpen
  \bibfield  {author} {\bibinfo {author} {\bibfnamefont {S.}~\bibnamefont {Utermann}},\ }\href@noop {} {\bibinfo {title} {Friction and diffusive light transport in a granular medium}} (\bibinfo {year} {2012})\BibitemShut {NoStop}%
\bibitem [{\citenamefont {Sperl}(2005)}]{sperl2005nearly}%
  \BibitemOpen
  \bibfield  {author} {\bibinfo {author} {\bibfnamefont {M.}~\bibnamefont {Sperl}},\ }\bibfield  {title} {\bibinfo {title} {Nearly logarithmic decay in the colloidal hard-sphere system},\ }\href@noop {} {\bibfield  {journal} {\bibinfo  {journal} {Phys. Rev. E}\ }\textbf {\bibinfo {volume} {71}},\ \bibinfo {pages} {060401} (\bibinfo {year} {2005})}\BibitemShut {NoStop}%
\bibitem [{\citenamefont {Sperl}\ \emph {et~al.}(2012)\citenamefont {Sperl}, \citenamefont {Kranz},\ and\ \citenamefont {Zippelius}}]{sperl2012single}%
  \BibitemOpen
  \bibfield  {author} {\bibinfo {author} {\bibfnamefont {M.}~\bibnamefont {Sperl}}, \bibinfo {author} {\bibfnamefont {W.~T.}\ \bibnamefont {Kranz}},\ and\ \bibinfo {author} {\bibfnamefont {A.}~\bibnamefont {Zippelius}},\ }\bibfield  {title} {\bibinfo {title} {Single-particle dynamics in dense granular fluids under driving},\ }\href@noop {} {\bibfield  {journal} {\bibinfo  {journal} {EPL}\ }\textbf {\bibinfo {volume} {98}},\ \bibinfo {pages} {28001} (\bibinfo {year} {2012})}\BibitemShut {NoStop}%
\bibitem [{\citenamefont {Charbonneau}\ \emph {et~al.}(2017)\citenamefont {Charbonneau}, \citenamefont {Kurchan}, \citenamefont {Parisi}, \citenamefont {Urbani},\ and\ \citenamefont {Zamponi}}]{charbonneau2017glass}%
  \BibitemOpen
  \bibfield  {author} {\bibinfo {author} {\bibfnamefont {P.}~\bibnamefont {Charbonneau}}, \bibinfo {author} {\bibfnamefont {J.}~\bibnamefont {Kurchan}}, \bibinfo {author} {\bibfnamefont {G.}~\bibnamefont {Parisi}}, \bibinfo {author} {\bibfnamefont {P.}~\bibnamefont {Urbani}},\ and\ \bibinfo {author} {\bibfnamefont {F.}~\bibnamefont {Zamponi}},\ }\bibfield  {title} {\bibinfo {title} {Glass and jamming transitions: From exact results to finite-dimensional descriptions},\ }\href@noop {} {\bibfield  {journal} {\bibinfo  {journal} {Annual Review of Condensed Matter Physics}\ }\textbf {\bibinfo {volume} {8}},\ \bibinfo {pages} {265} (\bibinfo {year} {2017})}\BibitemShut {NoStop}%
\bibitem [{\citenamefont {Coulais}\ \emph {et~al.}(2013)\citenamefont {Coulais}, \citenamefont {Candelier}, \citenamefont {Dauchot}, \citenamefont {Yu}, \citenamefont {Dong}, \citenamefont {Yang},\ and\ \citenamefont {Luding}}]{coulais2013Jam}%
  \BibitemOpen
  \bibfield  {author} {\bibinfo {author} {\bibfnamefont {C.}~\bibnamefont {Coulais}}, \bibinfo {author} {\bibfnamefont {R.}~\bibnamefont {Candelier}}, \bibinfo {author} {\bibfnamefont {O.}~\bibnamefont {Dauchot}}, \bibinfo {author} {\bibfnamefont {A.}~\bibnamefont {Yu}}, \bibinfo {author} {\bibfnamefont {K.}~\bibnamefont {Dong}}, \bibinfo {author} {\bibfnamefont {R.}~\bibnamefont {Yang}},\ and\ \bibinfo {author} {\bibfnamefont {S.}~\bibnamefont {Luding}},\ }\bibfield  {title} {\bibinfo {title} {{The Glass and Jamming transitions in dense granular matter}},\ }\href {https://doi.org/10.1063/1.4811862} {\bibfield  {journal} {\bibinfo  {journal} {{Powders and Grains 2013}}\ }\textbf {\bibinfo {volume} {1542}},\ \bibinfo {pages} {25} (\bibinfo {year} {2013})}\BibitemShut {NoStop}%
\bibitem [{\citenamefont {Behringer}(2015)}]{Behringer2015Jam}%
  \BibitemOpen
  \bibfield  {author} {\bibinfo {author} {\bibfnamefont {R.}~\bibnamefont {Behringer}},\ }\bibfield  {title} {\bibinfo {title} {Jamming in granular materials},\ }\href {https://doi.org/10.1016/j.crhy.2015.02.001} {\bibfield  {journal} {\bibinfo  {journal} {Comptes Rendus Physique}\ }\textbf {\bibinfo {volume} {16}} (\bibinfo {year} {2015})}\BibitemShut {NoStop}%
\bibitem [{\citenamefont {Silbert}\ \emph {et~al.}(2002)\citenamefont {Silbert}, \citenamefont {Erta\ifmmode~\mbox{\c{s}}\else \c{s}\fi{}}, \citenamefont {Grest}, \citenamefont {Halsey},\ and\ \citenamefont {Levine}}]{Silbert2002JamGlass}%
  \BibitemOpen
  \bibfield  {author} {\bibinfo {author} {\bibfnamefont {L.~E.}\ \bibnamefont {Silbert}}, \bibinfo {author} {\bibfnamefont {D.}~\bibnamefont {Erta\ifmmode~\mbox{\c{s}}\else \c{s}\fi{}}}, \bibinfo {author} {\bibfnamefont {G.~S.}\ \bibnamefont {Grest}}, \bibinfo {author} {\bibfnamefont {T.~C.}\ \bibnamefont {Halsey}},\ and\ \bibinfo {author} {\bibfnamefont {D.}~\bibnamefont {Levine}},\ }\bibfield  {title} {\bibinfo {title} {Analogies between granular jamming and the liquid-glass transition},\ }\href {https://doi.org/10.1103/PhysRevE.65.051307} {\bibfield  {journal} {\bibinfo  {journal} {Phys. Rev. E}\ }\textbf {\bibinfo {volume} {65}},\ \bibinfo {pages} {051307} (\bibinfo {year} {2002})}\BibitemShut {NoStop}%
\bibitem [{\citenamefont {Coniglio}\ and\ \citenamefont {Nicodemi}(2000)}]{Coniglio2000Glass}%
  \BibitemOpen
  \bibfield  {author} {\bibinfo {author} {\bibfnamefont {A.}~\bibnamefont {Coniglio}}\ and\ \bibinfo {author} {\bibfnamefont {M.}~\bibnamefont {Nicodemi}},\ }\bibfield  {title} {\bibinfo {title} {The jamming transition of granular media},\ }\href {https://doi.org/10.1088/0953-8984/12/29/331} {\bibfield  {journal} {\bibinfo  {journal} {Journal of Physics Condensed Matter}\ }\textbf {\bibinfo {volume} {12}} (\bibinfo {year} {2000})}\BibitemShut {NoStop}%
\bibitem [{\citenamefont {Goodrich}\ \emph {et~al.}(2015)\citenamefont {Goodrich}, \citenamefont {Liu},\ and\ \citenamefont {Sethna}}]{goodrich2015scaling}%
  \BibitemOpen
  \bibfield  {author} {\bibinfo {author} {\bibfnamefont {C.~P.}\ \bibnamefont {Goodrich}}, \bibinfo {author} {\bibfnamefont {A.~J.}\ \bibnamefont {Liu}},\ and\ \bibinfo {author} {\bibfnamefont {J.~P.}\ \bibnamefont {Sethna}},\ }\bibfield  {title} {\bibinfo {title} {Scaling theory for the jamming transition},\ }\href@noop {} {\bibfield  {journal} {\bibinfo  {journal} {arXiv preprint arXiv:1510.03469}\ } (\bibinfo {year} {2015})}\BibitemShut {NoStop}%
\bibitem [{\citenamefont {Schr{\"o}ter}(2017)}]{schroter2017local}%
  \BibitemOpen
  \bibfield  {author} {\bibinfo {author} {\bibfnamefont {M.}~\bibnamefont {Schr{\"o}ter}},\ }\bibfield  {title} {\bibinfo {title} {A local view on the role of friction and shape},\ }in\ \href@noop {} {\emph {\bibinfo {booktitle} {EPJ Web of Conferences}}},\ Vol.\ \bibinfo {volume} {140}\ (\bibinfo {organization} {EDP Sciences},\ \bibinfo {year} {2017})\ p.\ \bibinfo {pages} {01008}\BibitemShut {NoStop}%
\bibitem [{\citenamefont {D’Angelo}\ \emph {et~al.}(2022)\citenamefont {D’Angelo}, \citenamefont {Horb}, \citenamefont {Cowley}, \citenamefont {Sperl},\ and\ \citenamefont {Kranz}}]{Olfa2022Microgravity}%
  \BibitemOpen
  \bibfield  {author} {\bibinfo {author} {\bibfnamefont {O.}~\bibnamefont {D’Angelo}}, \bibinfo {author} {\bibfnamefont {A.}~\bibnamefont {Horb}}, \bibinfo {author} {\bibfnamefont {A.}~\bibnamefont {Cowley}}, \bibinfo {author} {\bibfnamefont {M.}~\bibnamefont {Sperl}},\ and\ \bibinfo {author} {\bibfnamefont {W.~T.}\ \bibnamefont {Kranz}},\ }\bibfield  {title} {\bibinfo {title} {Granular piston-probing in microgravity: powder compression, from densification to jamming},\ }\href@noop {} {\bibfield  {journal} {\bibinfo  {journal} {npj Microgravity}\ }\textbf {\bibinfo {volume} {8}},\ \bibinfo {pages} {48} (\bibinfo {year} {2022})}\BibitemShut {NoStop}%
\bibitem [{\citenamefont {Schröter}\ \emph {et~al.}(2007)\citenamefont {Schröter}, \citenamefont {Nägle}, \citenamefont {Radin},\ and\ \citenamefont {Swinney}}]{Schröter_2007}%
  \BibitemOpen
  \bibfield  {author} {\bibinfo {author} {\bibfnamefont {M.}~\bibnamefont {Schröter}}, \bibinfo {author} {\bibfnamefont {S.}~\bibnamefont {Nägle}}, \bibinfo {author} {\bibfnamefont {C.}~\bibnamefont {Radin}},\ and\ \bibinfo {author} {\bibfnamefont {H.~L.}\ \bibnamefont {Swinney}},\ }\bibfield  {title} {\bibinfo {title} {Phase transition in a static granular system},\ }\href {https://doi.org/10.1209/0295-5075/78/44004} {\bibfield  {journal} {\bibinfo  {journal} {Europhysics Letters}\ }\textbf {\bibinfo {volume} {78}},\ \bibinfo {pages} {44004} (\bibinfo {year} {2007})}\BibitemShut {NoStop}%
\bibitem [{\citenamefont {Mandal}\ \emph {et~al.}(2020)\citenamefont {Mandal}, \citenamefont {Nicolas},\ and\ \citenamefont {Pouliquen}}]{Madel2021}%
  \BibitemOpen
  \bibfield  {author} {\bibinfo {author} {\bibfnamefont {S.}~\bibnamefont {Mandal}}, \bibinfo {author} {\bibfnamefont {M.}~\bibnamefont {Nicolas}},\ and\ \bibinfo {author} {\bibfnamefont {O.}~\bibnamefont {Pouliquen}},\ }\bibfield  {title} {\bibinfo {title} {Insights into the rheology of cohesive granular media},\ }\href {https://doi.org/10.1073/pnas.1921778117} {\bibfield  {journal} {\bibinfo  {journal} {Proceedings of the National Academy of Sciences}\ }\textbf {\bibinfo {volume} {117}},\ \bibinfo {pages} {8366} (\bibinfo {year} {2020})},\ \Eprint {https://arxiv.org/abs/https://www.pnas.org/doi/pdf/10.1073/pnas.1921778117} {https://www.pnas.org/doi/pdf/10.1073/pnas.1921778117} \BibitemShut {NoStop}%
\bibitem [{\citenamefont {Papadopoulos}\ \emph {et~al.}(2017)\citenamefont {Papadopoulos}, \citenamefont {Porter}, \citenamefont {Daniels},\ and\ \citenamefont {Bassett}}]{Papa2007}%
  \BibitemOpen
  \bibfield  {author} {\bibinfo {author} {\bibfnamefont {L.}~\bibnamefont {Papadopoulos}}, \bibinfo {author} {\bibfnamefont {M.}~\bibnamefont {Porter}}, \bibinfo {author} {\bibfnamefont {K.}~\bibnamefont {Daniels}},\ and\ \bibinfo {author} {\bibfnamefont {D.}~\bibnamefont {Bassett}},\ }\bibfield  {title} {\bibinfo {title} {Network analysis of particles and grains},\ }\href {https://doi.org/10.1093/COMNET/CNY005} {\bibfield  {journal} {\bibinfo  {journal} {Journal of Complex Networks}\ }\textbf {\bibinfo {volume} {6}} (\bibinfo {year} {2017})}\BibitemShut {NoStop}%
\end{thebibliography}%
\balancecolsandclearpage

\appendix
\section*{Appendix: Modeling the Agitation}

\begin{figure}[h]
    \centering
    \includegraphics[width=0.9\linewidth]{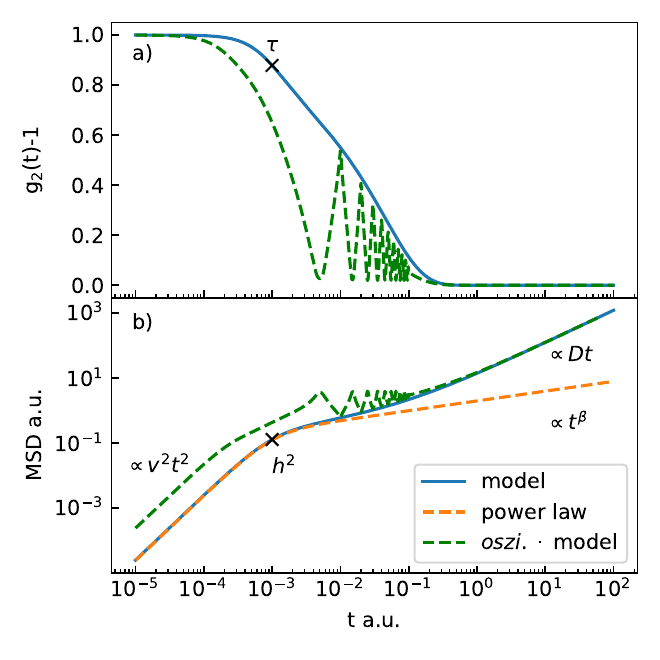}
    \caption{Schematic illustration of a) correlations function and b) mean squared displacement of the model function with (green) and without (blue) agitating oscillation as described in the text.}
    \label{fig1b}
\end{figure}

We assume that the measured intensity correlation function $g_{2,\textbf{m}}$ is a convolution, weighted with $C$, between the oscillation $O(t)$ with the intensity correlation function $g_2(t)$, and the pure $g_2(t)$ correlation function.
\begin{equation}
g_{2,\textbf{m}} = g_2(t) (C O(t) + (1-C))   
\label{eq:oszimodel}
\end{equation}
The oscillation $O(t)$ of the agitation is given by 
\begin{equation}
O(t) = \frac{1}{T} \int_0^T \exp( -\kappa^2 A^2 P(\omega(t + t')) - P(\omega t') )^2 dt'
\label{eq:oszi}
\end{equation}
and the correlation of the periodic rectangular agitation function $P(t)$ is modelled by the profile shown in Fig. \ref{fig1b} which is obtained by measuring the applied voltage profile. We measure $g_2(t)$ and assume that the field correlation function $g_1(t)$

To construct $g_1(t)$ we use a model function for the MSD with power laws for ballistic, sub-diffusion, and diffusion behavior
\begin{equation}
    \label{eq:JanFit}
    \langle \Delta r^2 \rangle = \frac{h^2}{ (\frac{\tau}{t})^2+ (\frac{\tau}{t})^{\beta}}   +   6Dt+6D \tau \left(e^{-\frac{t}{\tau}}-1\right)
\end{equation}

whit ballistic time constant $\tau$, diffusion coefficient $D$, localization length $h$ and power law exponent $\beta$ (illustrated in Fig. \ref{fig1b}) see also \cite{Kunzner2025VC}.\textbf{}

\end{document}